\documentclass[preprint,3p,times,12pt]{elsarticle}

\usepackage{cleveref}
\crefname{figure}{figure}{figures}
\usepackage{caption}
\usepackage{subcaption}
\usepackage[UKenglish]{babel}
\usepackage{alltt}
\usepackage{epstopdf}
\usepackage{setspace}
\usepackage{float}
\usepackage{amsmath}
\usepackage{dcolumn}% Align table columns on decimal point
\usepackage[perpage]{footmisc}
\usepackage{amsmath,amsfonts,amssymb,wasysym,ifsym}
\usepackage{color}
\usepackage{verbatim}
\usepackage{upgreek}
\usepackage{hyperref}
\usepackage{bm}% bold math

\usepackage{graphicx}
\usepackage{lineno}

\newcommand \lith{$^{6}$Li}

\newcommand \LiF{\lith F:ZnS(Ag)}

\begin{document}

\begin{frontmatter}



\title{Pulse Shape Discrimination and Exploration of Scintillation Signals Using Convolutional Neural Networks}

%This file is included in main document by using "\input{SoLid_author_list}"

%Something to allow multiple use of same footnote
\makeatletter
\makeatother

%Institutes: with short name as new command
%List of authors and their corresponding institute
\author{J.~Griffiths \fnref{add1}}
\fntext[add1]{Now at: University of Cambridge, Cavendish Laboratory, Cambridge, CB3 0HE, United Kingdom}

\author{S.~Kleinegesse \fnref{add2}}
\fntext[add2]{Now at: University of Edinburgh, School of Informatics, Edinburgh, EH8 9AB, United Kingdom}

\author{D.~Saunders}
\author{R.~Taylor}

\author{A.~Vacheret \corref{cor1}} 
\ead{antonin.vacheret@imperial.ac.uk}
\cortext[cor1]{Corresponding author}
\address{Imperial College London, Department of Physics, London, SW7 2AZ, United Kingdom}

%\begin{collab}
%\centering (YYY Collaboration)
%\end{collab}

\date{\today}

\begin{abstract}
We demonstrate the use of a convolutional neural network to perform neutron-gamma pulse shape discrimination, where the only inputs to the network are the raw digitised SiPM signals from a dual scintillator detector element made of \LiF~scintillator and PVT plastic. A realistic labelled dataset was created to train the network by exposing the detector to an AmBe source, and a data-driven method utilsing a separate PMT was used to assign labels to the recorded signals. This approach is compared to the charge integration and continuous wavelet transform methods and is found to provide superior levels of discrimination, achieving an AUC of $0.995\pm0.003$. We find that the neural network is capable of extracting interpretable features directly from the raw data. In addition, by visualising the high-dimensional representations of the network with the t-SNE algorithm, we discover that not only is this method robust to minor mislabeling of the training dataset but that it is possible to identify an underlying substructure within the signals that goes beyond the original labelling. This technique could be utilised to explore and cluster complex, raw detector data in a novel way that may reveal more insights than standard analysis methods.
\end{abstract}


\end{frontmatter}

%\linenumbers


\section{Introduction}
% !TEX root = ../ms.tex
\ifpdf
    \graphicspath{{Sections/}}
\else
    \graphicspath{{Sections/}}
\fi
Efficient pulse shape discrimination (PSD) techniques are required in many applications, from radiation measurements to particle physics experiments. PSD is a typical classification problem whereby digitised waveforms can be discriminated based on their time and energy features. Such features or characteristics are usually dictated by the underlying physical process that occurs in the detection medium. For example, in scintillation signals, a common characteristic used for PSD is the decay time of the detected scintillation light, which is dictated by the atomic or molecular structure of radiative states. 

In this paper we focus on one specific application of PSD using \LiF\ phosphor screens similar to those used in the SoLid reactor neutrino experiment \cite{SoLid} and in other specific neutron detection applications where high discrimination is required. The neutron capture reaction on $\mathrm{^{6}Li}$ produces highly ionising particles that excite the ZnS(Ag) scintillator:  
\begin{equation}
\label{eq:ibd}
    n + \mathrm{^{6}Li} \rightarrow \alpha + \mathrm{^{3}_{1}H}\ (4.8~\text{MeV})
\end{equation}
Thin sheets of \LiF~ provide high neutron detection efficiency due to the large \lith\ neutron capture cross section and the high scintillation yield of ZnS(Ag). Strong gamma-ray background rejection can be achieved as a result of the large difference between fast and slow ZnS(Ag) scintillation components for electron and nuclear interactions, respectively. 
As our focus here is the classification of those two type of signals, we make the distinction between these scintillation responses by defining as {\it electron scintillation} (ES) signals, the interactions from gamma-rays and other charged particles (electron, positrons, muons, etc.) and {\it nuclear scintillation} (NS) signals as those produced by nuclei such as protons, alpha particles or heavier ions. The $\mathrm{^{6}Li}$ neutron capture reaction that produces a tritium and alpha therefore produces a distinctive NS signal in this detector medium.    

Traditional pulse-shape discrimination (PSD) techniques that utilise time domain information such as charge-integration \cite{LiFZnS.PSD} are popular for their robustness and reasonable performance. Other methods based on frequency information \cite{wavelet} can achieve superior performance by exploiting better the information available, but are more computationally expensive. The performance of these approaches is dependent on a number of factors that include the choice of scintillator, experimental requirements and read out technique. They also tend to have limitations with real data that exhibit a number of additional features such as pile up or background interactions. 

Neural networks have the potential to be superior classifiers provided they are trained with realistic data. In particular, convolutional neural networks (CNN), are well suited to raw data that have high local correlations such as waveform signals. They are very successful in computational vision tasks and can reach high performance with limited datasets. Once trained, obtaining a classification output from a CNN is also very fast.

In this work, we present the results of using a CNN to discriminate between ES and NS signals using the raw scintillation pulse from a single PVT cube with \LiF. 
First, we describe the experimental set up that was used to collect a labelled dataset for the training of the network. We then present the details of the CNN architecture as well as two other common PSD algorithms, in order to provide a set of benchmarks that are compared to the results of the CNN. Finally we investigate and discuss the features learned by the network, demonstrating that they are interpretable and can be utilised to identifty a clear substructure within the ES and NS signals.

\section{Experimental Dataset}
\label{exp}

\begin{figure}[]
\begin{center}
\includegraphics[width=0.7\textwidth]{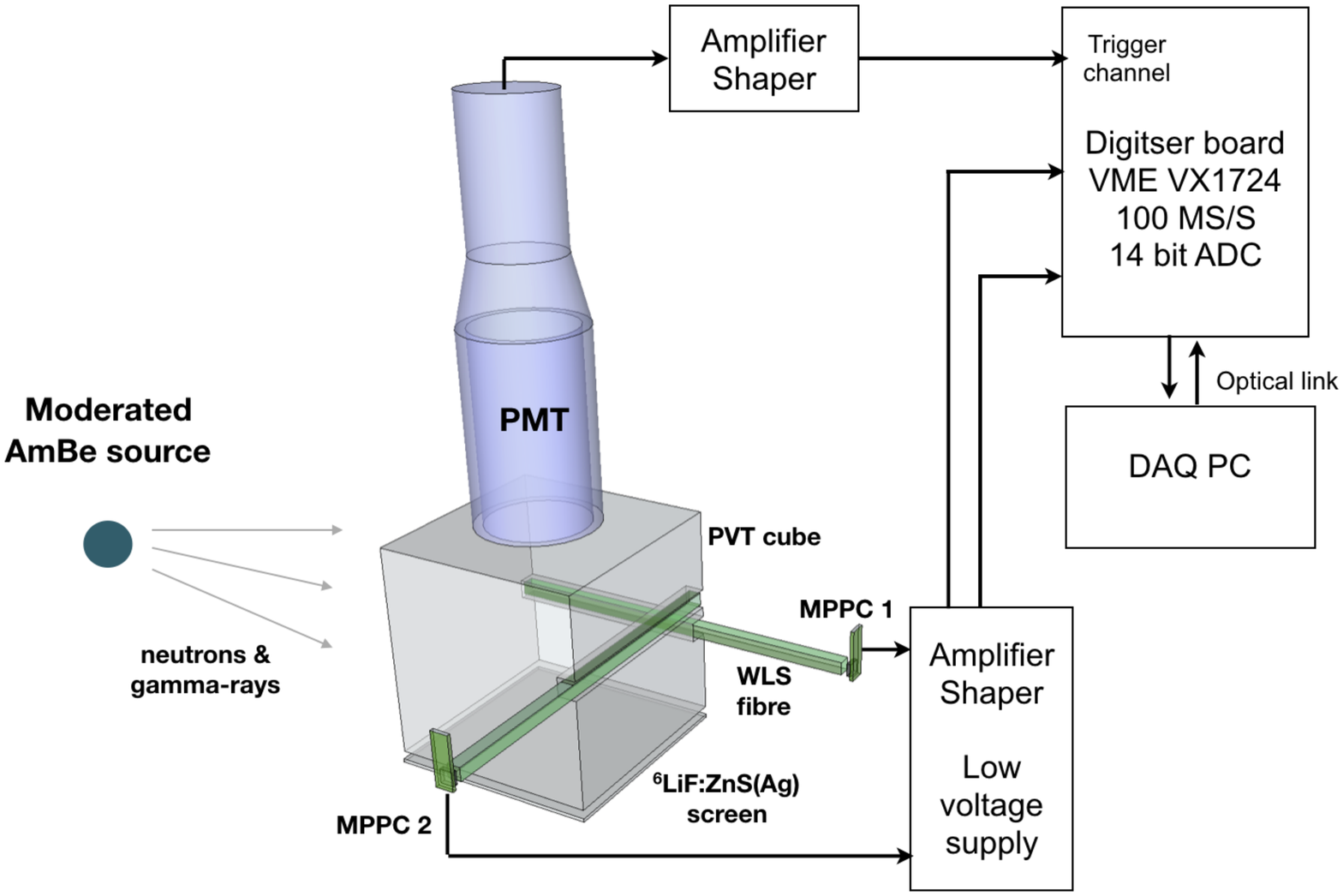}
\caption{\label{fig:schematicdataset} Schematic of the experimental set-up used to record scintillation signals from the PVT cube.}
\end{center}
\end{figure}

In this work, supervised learning was used to train the CNN and therefore a labelled dataset of ES and NS signals was required. Instead of using a set of Monte-Carlo generated signals which may contain unrealistic features, a dataset consisting of real scintillation signals was collected. A schematic of the experimental setup used to collect this data is shown on Fig.~\ref{fig:schematicdataset}. The detector element consists of a neutron sensitive \LiF~screen coupled to a polyvinyltoluene (PVT) scintillating cube, which is only sensitive to ES signals. The detector element was exposed to an AmBe source that emits neutrons and gamma-rays over a wide range of energies, and therefore provides a realistic range of signals for the training set. A small fraction of the dataset includes signals from background interactions such as muons and natural radioactivity and the reasonable activity of the source also results in a fraction of pile up pulses. Scintillation light from both scintillators was collected by wavelength shifting fibres placed in grooves on the side of the cube and registered by two Hamamatsu MPPC S12572-050P silicon photomulitpliers (SiPMs) attached to one end of each fibre. To enhance the reflection of light inside the cube and collection of that light by the fibres, the detector element was wrapped in Tyvek. 

A Philips XP1911/UVPA photomultiplier tube (PMT) was placed on top of a hole cut into the Tyvek, and used to detect a large fraction of the blue scintillation light that is not collected by the fibres. The greater collection of emitted light in the PMT was intended to provide strong discrimination of NS and ES signal so that labels can be confidently assigned to the corresponding signals seen by the SiPMs. The data from both the PMT and SiPMs was digitised using a CAEN VX1724 digitiser card at 100 MS/s with a 14 bit range, and the larger PMT response was used to trigger the acquisition of the SiPMs signals. To ensure that the majority of the slow ZnS scintillation component was acquired, the maximum length of each waveform was set to be 1000 samples. 

\begin{figure}[]
\begin{center}
\includegraphics[width=0.6\textwidth]{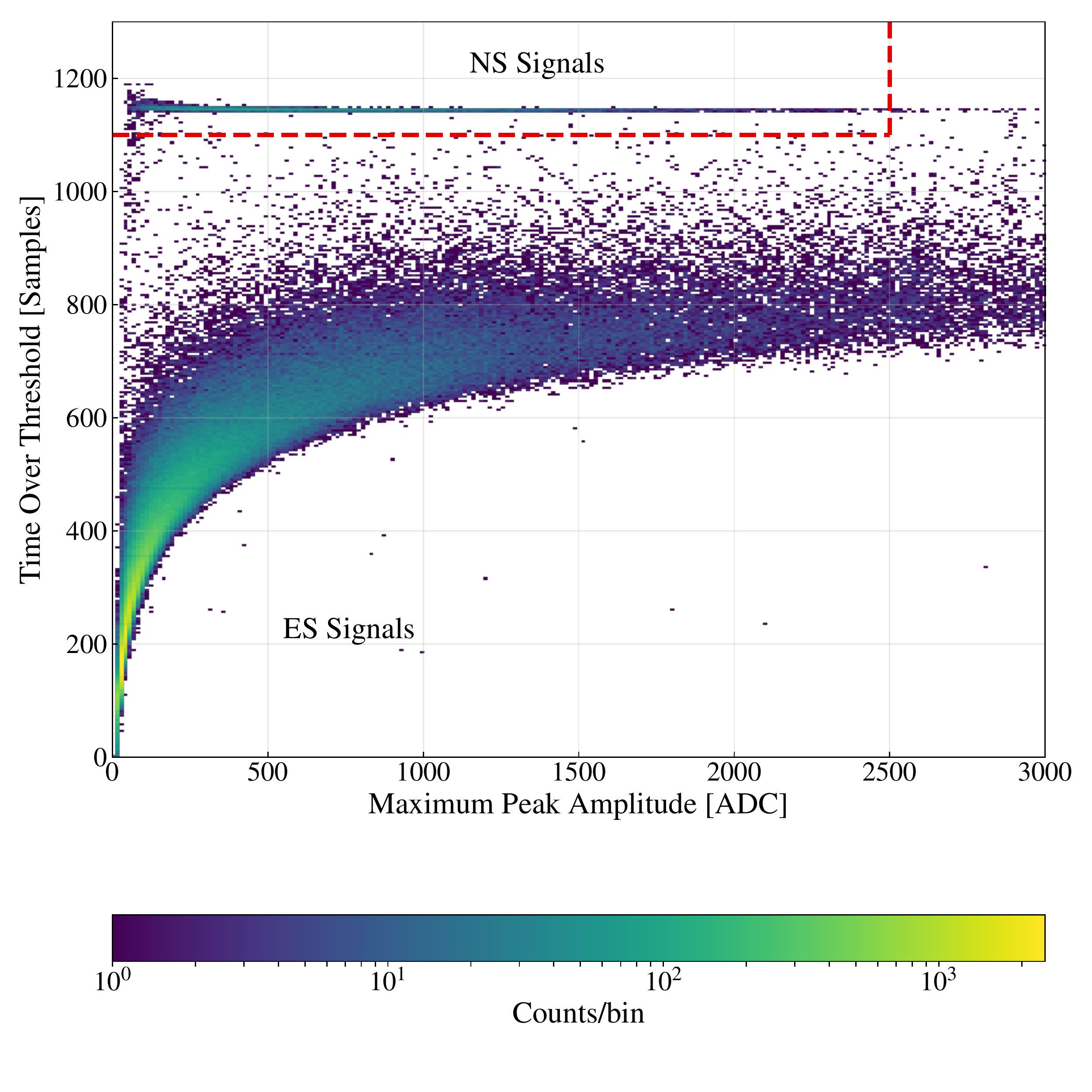}
\caption{\label{fig:psddataset} Distribution of PMT signals based on their time ove threshold value versus the amplitude of the maximum peak captured in the waveform. The dashed red line shows the selection cut used for labelling the data.}
\end{center}
\end{figure}

\begin{figure}[]
\begin{center}
\includegraphics[width=0.6\textwidth]{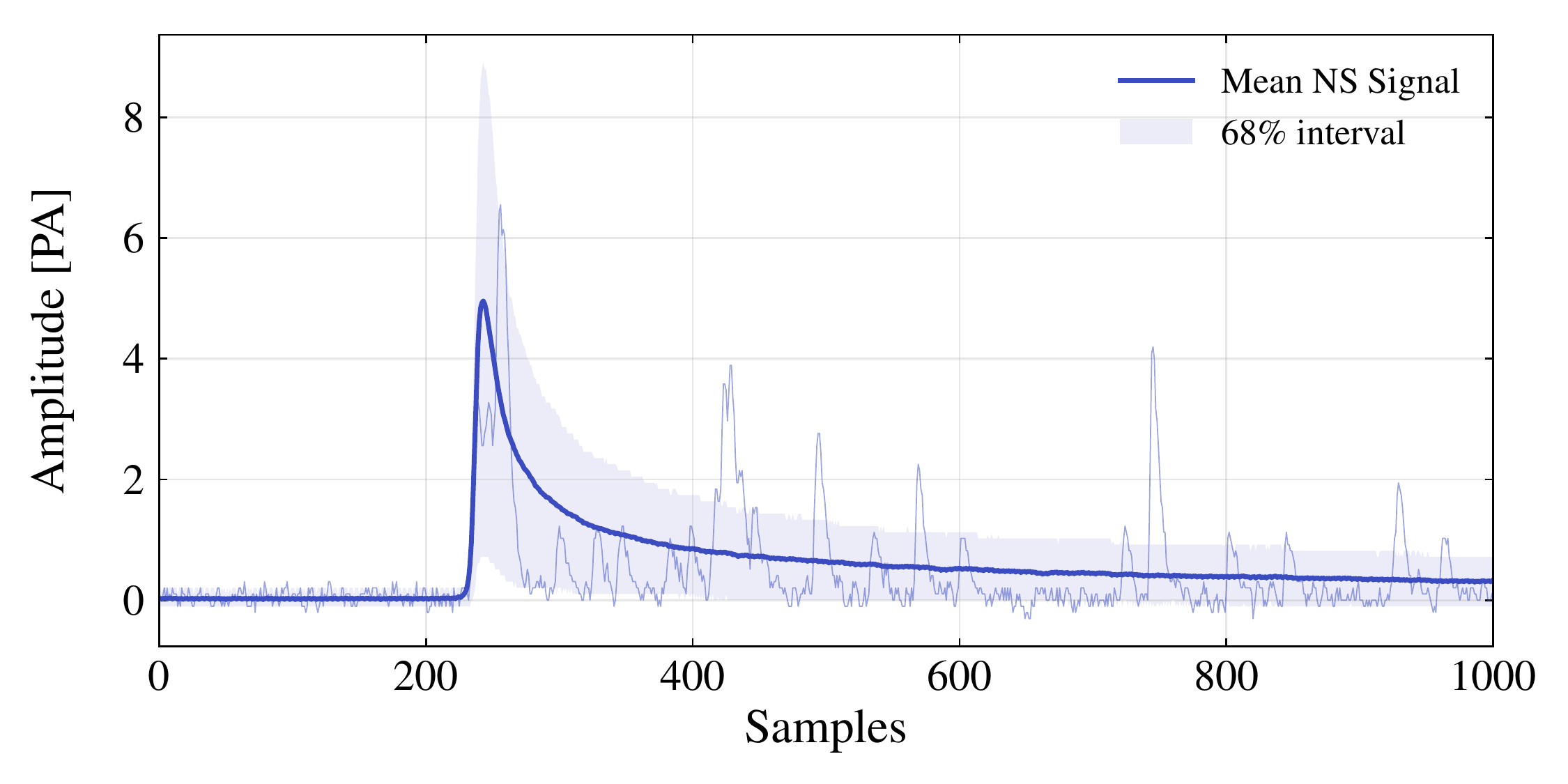}
%\caption{\label{fig:ESwf} Raw waverform acquired for neutron candidate (top) and gamma-ray signals (bottom).}
\includegraphics[width=0.6\textwidth]{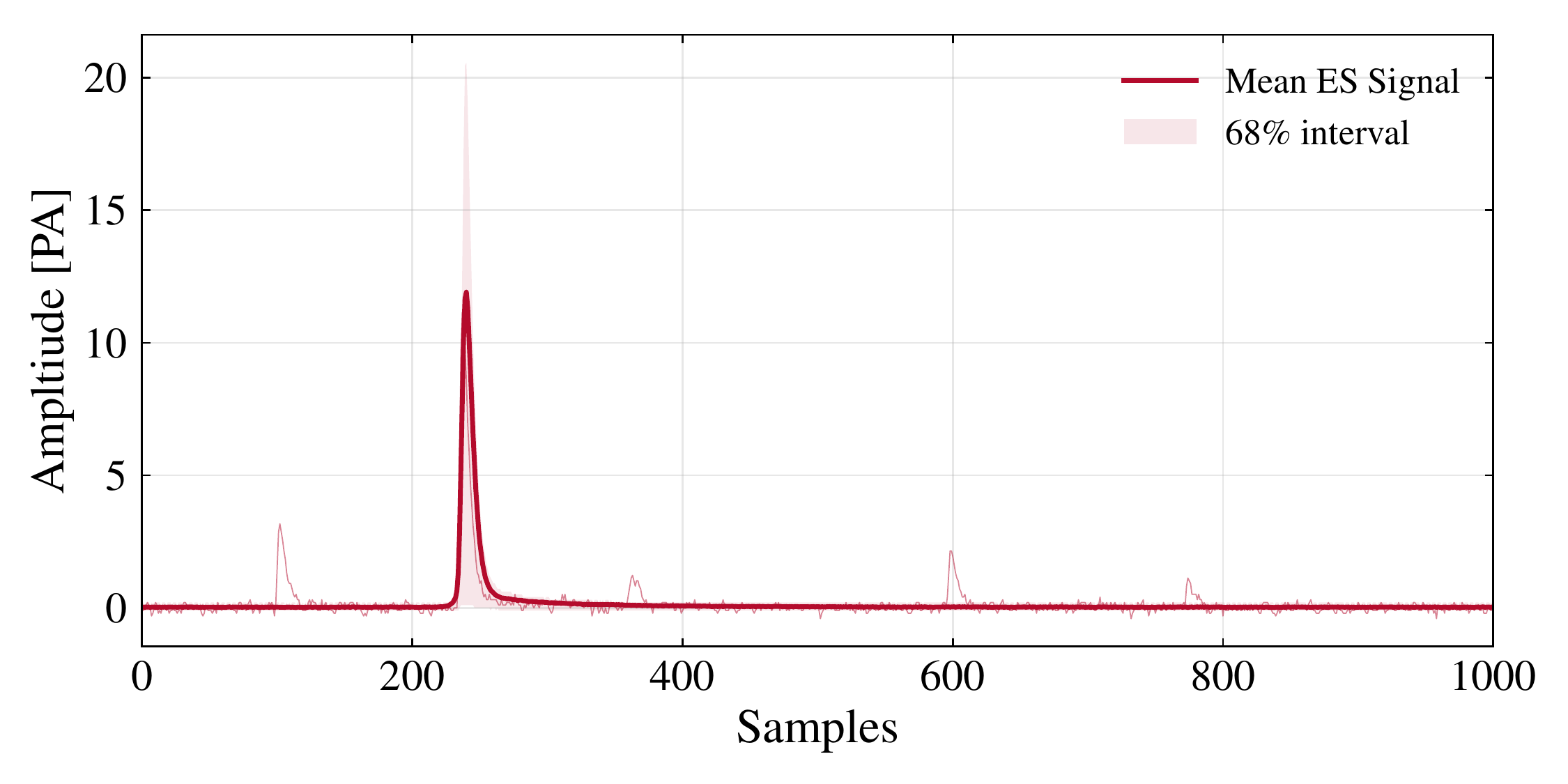}
    \caption{\label{fig:wf} Average NS (top) and ES (bottom) digitised signals from one of the SiPMs. The shaded bands represent the central 68$\%$ interval of the signal distribution. An example waveform is also shown for each signal type.}
\end{center}
\end{figure}

Labels were then assigned to the NS and ES signals using a simple PSD method based on the time-over-threshold (TOT) and maximum peak amplitude (MPA) of the PMT signals, which provides very good results for discriminating events at a low threshold. Fig.~\ref{fig:psddataset} shows the PSD parameter value as a function of signal amplitude for each of the PMT waveforms. As a result of the long shaping of the PMT pulse, NS events have a much longer decay time and are above threshold for the majority of the acquired time period. Signals with TOT $\geq$ 1100 samples and MPA $\leq$ 2500 ADC are labelled as NS and those outside of this range are labelled as ES. The identification of NS signals is very robust, and we estimated that $~1\%$ of this sample has contamination from other low amplitude events. Typical waveforms for neutrons (NS) and other signals such as gamma-rays (ES) are shown in Fig.~\ref{fig:wf}.

\section{Pulse Shape Discrimination}
Pulse shape discrimination techniques utilise the dominant shape and amplitude features of digitised signals by selecting a subset of the total information and compressing this into a reduced quantity. An alternative approach that can take advantage of all the information within a signal is to directly use the digitised signals as an input to a suitable machine learning algorithm, which is trained to assign labels to individual signals. This has been shown to provide superior performance in specific applications \cite{Flores:2016dfj}\cite{6551092}. An ideal machine learning algorithm for pulse shape discrimination are convolutional neural networks (CNN) \cite{deeplearning} as they have the ability to efficiently extract and combine locally correlated features directly from raw data. As a result they have become the established technique for complex image recognition problems and in recent work have been applied in high-energy physics data analysis where they have been shown to provide comparable or improved performance in tasks such as background rejection \cite{Komiske2017} \cite{next} \cite{panda}, event reconstruction \cite{Delaquis}, and event classification \cite{DayaBay} \cite{NOVA} \cite{miniboone} when trained on the raw detector outputs. In this section we introduce the optimised CNN architecture used for classifying ES and NS signals, as well as two commonly used PSD techniques: the charge integration and continuous wavelet transform methods whose results are used as a benchmark to compare against. 

\subsection{Charge Integration}
\label{CI}
The charge integration (CI) method~\cite{LiFZnS.PSD} is a a robust and easily interpretable method of pulse shape discrimination. It uses the ratio of pulse area contained in the tail of a pulse to that contained within a short, initial time period. For a signal $f(t)$, a short integration window, $Q_{short} = \int^{t_{short}}_0 f(t) dt$, and long integration window $Q_{long} = \int^{t_{long}}_0 f(t) dt$ are defined and the quantity
\begin{equation}
CI = \frac{Q_{long} - Q_{short}}{Q_{long}}
\end{equation}
can be used to discriminate between pulse types. 
The optimal values of the parameters $t_{short}$ and $t_{long}$ were found to be 245~ns and 497~ns, respectively.

\subsection{Continuous Wavelet Transform}
\label{CWT}
The continuous wavelet transform (CWT) \cite{wavelet} provides a powerful method of pulse shape discrimination that is capable of utilising information from both the time and frequency domains of a signal. The continuous wavelet transform of a signal $f(t)$ at a scale $a$ and shift $b$ is defined as
\begin{equation}
W_f(a,b) = \frac{1}{\sqrt{a}} \int_{-\infty}^{\infty} f(t) \psi \left( \frac{t-b}{a} \right) dt,
\end{equation}
which can be interpreted as the convolution of the signal with a series of scaled and shifted wavelets $\psi(t)$. The energy density of the wavelet transform at a scale $a$ is defined as
\begin{equation}
P(a) = \frac{1}{1+n_b} \sum_{j=0}^{n_b} |W_f(a,b_j)|^2,
\end{equation}
which depends heavily on the shape of the signal. Two different scales, $a_1$ and $a_2$, can therefore be chosen to discriminate between different signal shapes using the variables $f_1 = P(a_1)$ and $f_2 = P(a_1) / P(a_2)$. A hyperplane in the ($f_1$, $f_2$) space can then be used to select different signal shapes. Using the Ricker wavelet, the optimal values of $a_1$ and $a_2$ were found to be 1 and 900, respectively.

\subsection{Convolutional Neural Network}
\label{CNN}
\begin{figure*}[h]
\centering
\includegraphics[width=\textwidth]{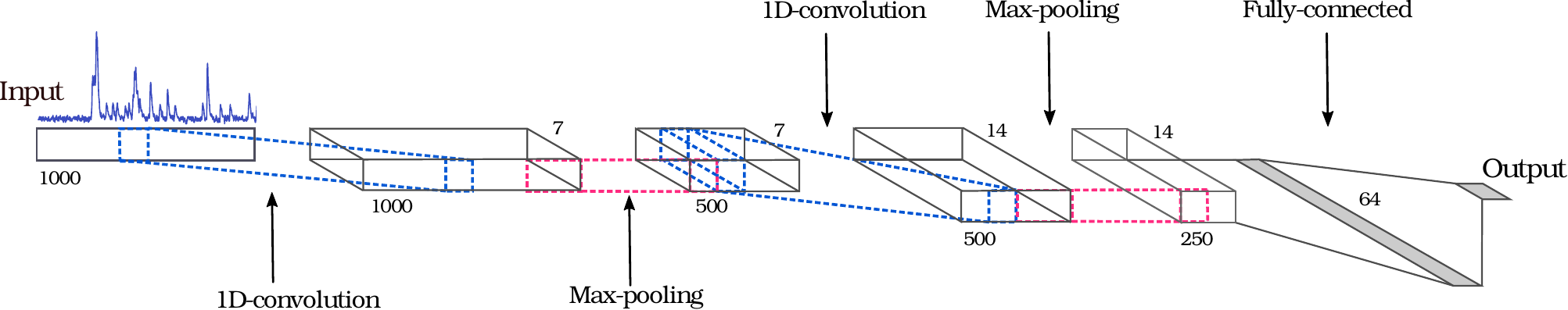}
\caption{\label{fig:architecture} Schematic of the CNN architecture used in this work. The inputs to the network are individual signals of length 1000 samples. Two successive convolutional and pooling layers extract features from the signal which are combined in the fully-connected layer. The output layer consists of a single neuron with the softmax activation, which represents the probability of a signal being NS.}
\end{figure*}
CNNs are an extension to feed-forward neural networks that are capable of extracting locally correlated features from a multi-dimensional input. This is achieved through the use of \textit{convolutional layers} and \textit{pooling layers}. For a one-dimensional input of length $k$, a convolutional layer is specified by a set of $\alpha$ filters, each with a width $m$ and a set of learnable weights $w_i$ ($i=1...m$). This layer transforms an input $f$ with $c$ separate channels each of length $k$ through the operation
\begin{equation}
F^\alpha_i = g \left (\sum_{\beta=1}^c \sum_{m=1}^k h^{\alpha \beta}_i f^\beta_{i+m} \right ),
\end{equation}
where $f^\beta$ is the $\beta$ channel of the input, $h^{\alpha \beta}_i$ are the learnable weights of the filter of the $\alpha$ output channel applied to the $\beta$ channel of the input, and $g$ is a non-linear activation function. The resulting outputs $F^\alpha_i$ are the \textit{feature maps}, which can be interpreted as a non-linear representation of the input that consists of features extracted by the filters. The filters are learnt during a supervised learning process that minimises a loss function, which for ES/NS discrimination is the standard binary logistic loss.
The pooling layers perform dimensionality reduction of the feature maps, and therefore reduce the total number of learnable parameters of the network. \textit{Max-pooling}, defined by taking the maximum value in a small region of the input, is used in this work as it provides a degree of translation invariance. 
Common CNN architectures use successive convolutional and pooling layers to produce multiple abstract representations of the input, which can then be combined to perform classification. Further explanation of CNNs can be found in \cite{nn_review}.

In this work, a CNN architecture was developed that classifies the raw, digitised signals obtained from the experimental set-up introduced in section 2 as either ES or NS signals. For a fair comparison with conventional PSD techniques only one of the SiPM signals was used. A schematic of the complete architecture is given in Figure~\ref{fig:architecture}, with the specific parameters of the convolutional and pooling layers given in Table \ref{tab:params}. The CNN transforms each signal through two successive convolution and pooling layers (window size of 2), where the non-linear ReLU activation function was used in both convolutional layers. The resulting feature maps are fed into a fully-connected layer with 64 neurons, that each have the ReLU activation. The output of the CNN consists of a single neuron with the softmax activation, and can therefore be be interpreted as the probability of the signal being NS. To classify signals, a threshold must be determined above (below) which all signals will be classified as NS (ES). Keras \cite{chollet2015keras} was used with the Tensorflow \cite{tensorflow2015-whitepaper} backend to train the CNN. The weights of the network were optimised by minimising the cross-entropy loss on a training sample of 12 000 signals with an equal proportion of ES and NS signals. The network was trained for a total of 50 epochs using the Adam optimsier \cite{adam} with a batch size of 256 samples and initial learning rate of 0.001. 

\begin{table}[]
     \centering
     \begin{tabular}{ l l l l l }
     \hline
     Layer & \# Channels & Filter Size  \\ \hline
     Conv & 7 & 10  \\ %\hline
     Max-pool & -- & 2  \\ 
     Conv & 14 & 10  \\ %\hline
     Max-pool & -- & 2  \\ %\hline

     \hline
     \end{tabular}
     \caption{Parameters of the convolutional and pooling layers of the CNN architecture. }
     \label{tab:params}
 \end{table}

\section{Results}
\begin{figure}[]
\begin{center}
\includegraphics[width=0.8\textwidth]{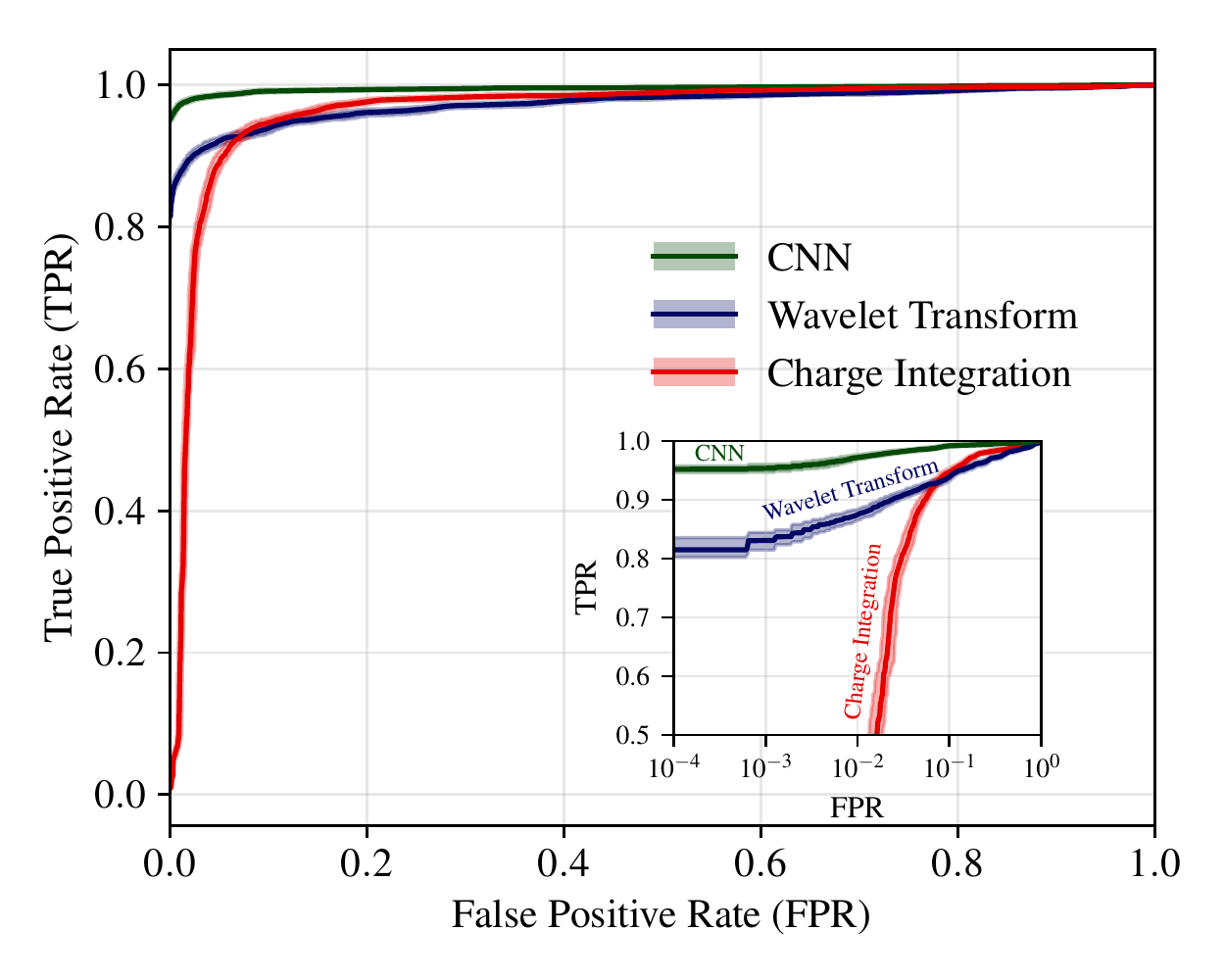}
\caption{\label{fig:roc_supervised} ROC curves of the CNN (green), CWT (blue) and CI (red) methods. The shaded bands represent the $1\sigma$ deviation on the true positive rate obtained by repeated sampling of the test set.}
\end{center}
\end{figure}

The performance of the CNN as well as the CI and CWT methods was evaluated on a seperate test sample of 3~000 signals. Figure~\ref{fig:roc_supervised} compares the ROC curves of the CNN to the CI and CWT methods on this sample, with the AUC metrics given in Table \ref{tab:auc}. Errors on the ROC curves and AUC metric are estimated by repeatedly evaluating the performance on random samples of the test set. The CNN is able to achieve a significantly higher level of ES/NS discrimination compared to CWT and CI methods, with an AUC of 0.995 $\pm$ 0.003.
\begin{table}[]
     \centering
     \begin{tabular}{ l  l }
     \hline
     Method & AUC \quad ($\pm 1\sigma$)  \\ \hline
     CNN &  $0.995 \quad (\pm 0.003)$ \\ %\hline
     CWT &  $0.974 \quad (\pm 0.003)$ \\ %\hline
     CI  &  $0.964 \quad (\pm 0.004)$ \\
     \hline
     \end{tabular}
     \caption{Area under the curve (AUC) values for each of the PSD methods considered in this work, with $1\sigma$ uncertainties given in brackets.}
     \label{tab:auc}
 \end{table}

\section{Feature Interpretation}

\label{Interp}
\begin{figure*}[]
\centering
\begin{subfigure}[]{0.45\textwidth}
\includegraphics[width=\textwidth]{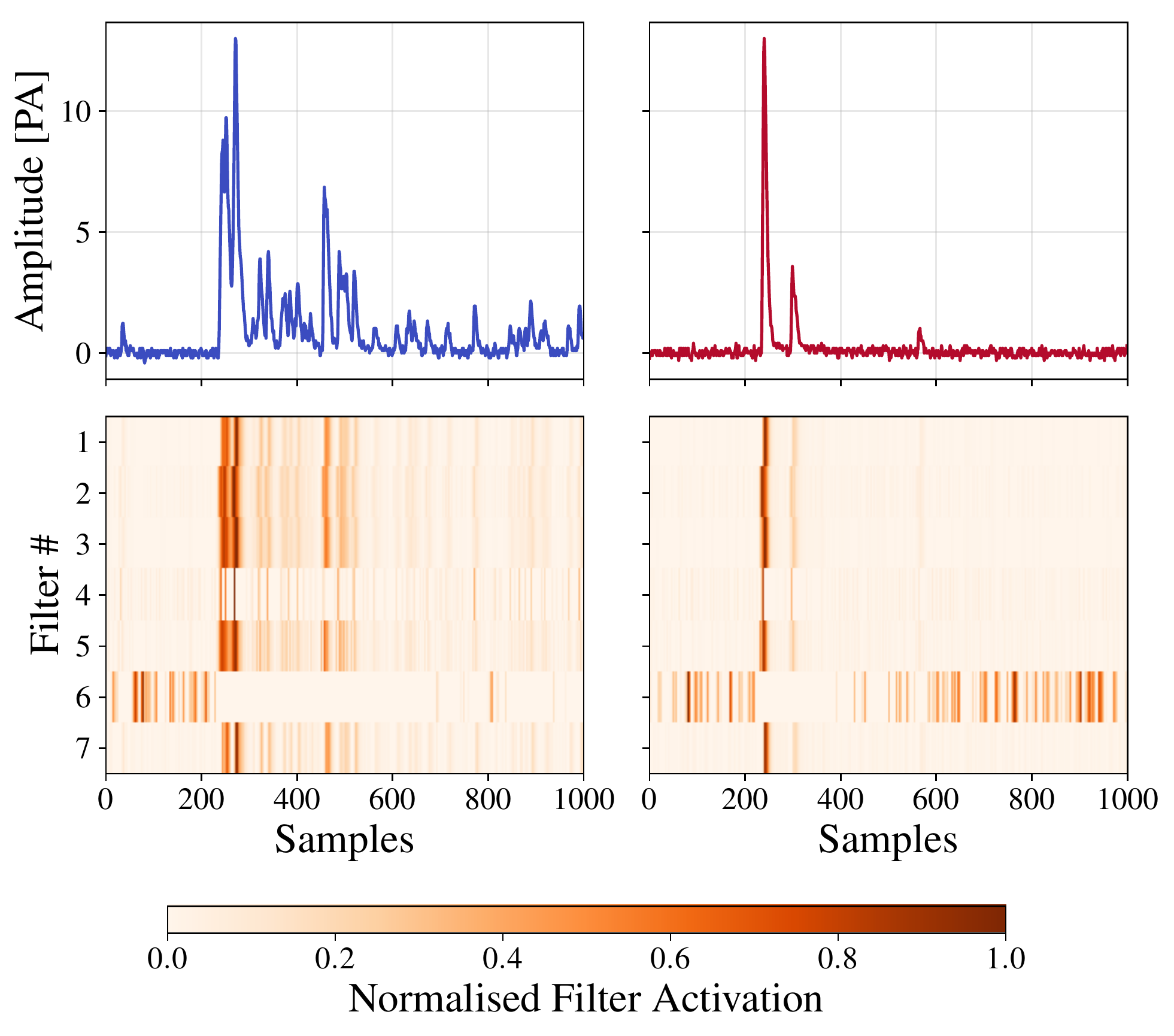}
\end{subfigure}
\begin{subfigure}[]{0.455\textwidth}
\includegraphics[width=\textwidth]{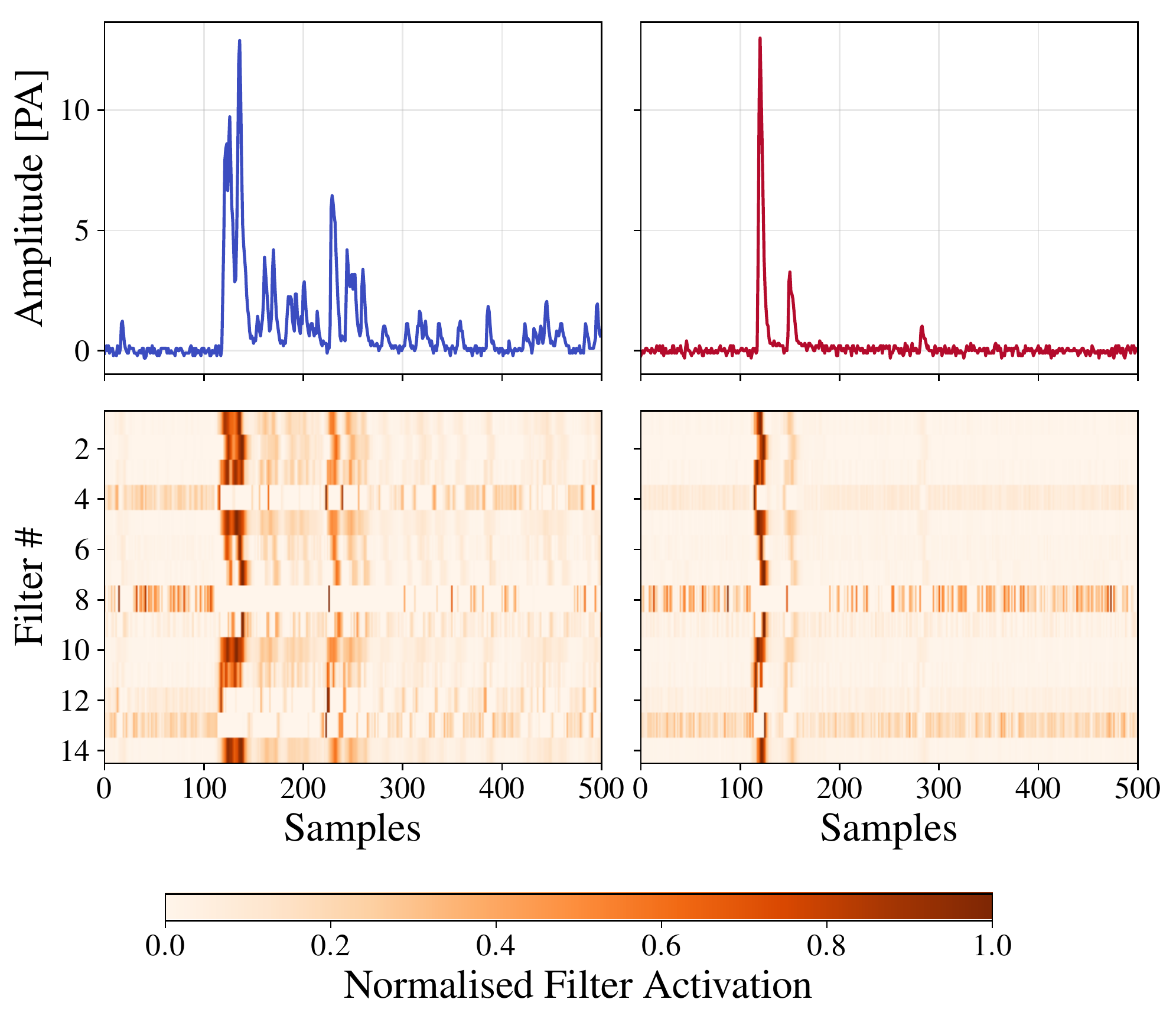}
\end{subfigure}
\caption{\label{fig:cnn_filters} Visualisation of the normalised filter outputs of the first (left) and second (right) convolutional layers for representative NS (blue) and ES (red) signals. In the right plot, the input signal has been downsampled to show the correlation with the second convolutional layer.}
\end{figure*}
The superior performance of the CNN is due to its ability to efficiently extract and combine the features relevent for discrimination, directly from the raw signal. To understand these features, the normalised feature maps of the two convolutional layers are visualised in Fig.~\ref{fig:cnn_filters} for representative NS and ES signals. It can be seen that in the first convolutional layer, most of the filters activate at the peaks of the signals and it can be see that filter 4 activates only at the maximum of each peak. This allows the CNN to encode both the number of peaks in a signal along with their relative amplitudes. Filter 6 appears to activate when the signals are below some threshold, which allows the CNN to extract quantities similar to time-over-threshold. It can be seen that in the second convolutional layer the feature maps become more complex, with multiple features of the first convolutional layer combined. For example, filter 4 appears to activate both on low ampltitude regions of the signal as well as at the steeply rising edges of high amplitude pulses.   

\begin{figure}[]
\begin{center}
\includegraphics[width=0.9\textwidth]{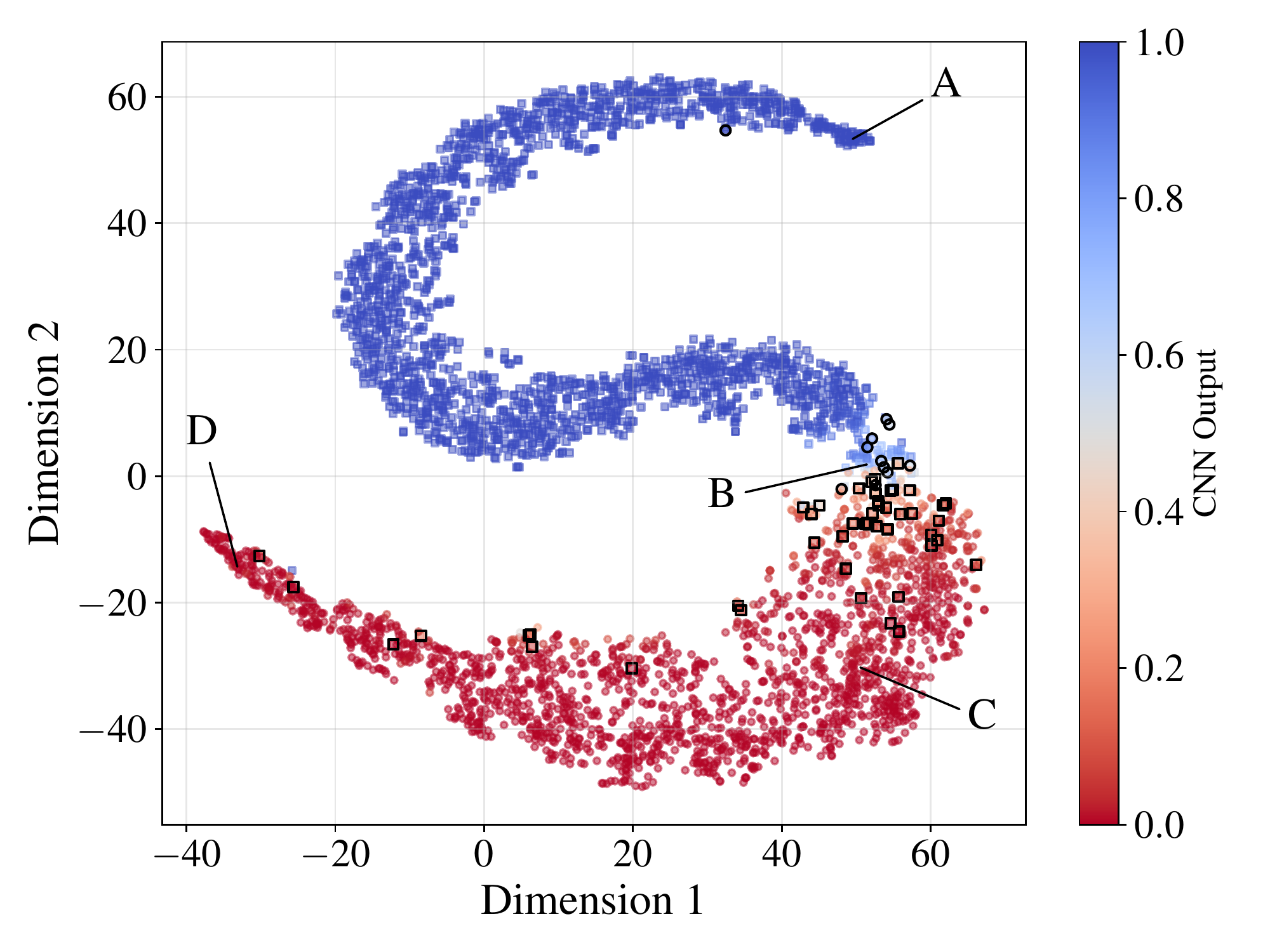}
\caption{\label{fig:tsne} t-SNE embedding of the fully-connected layer of the CNN. Each point in this two-dimensional space represents a single scintillation signal. ES signals are represented by circles and NS signals are represented by squares. The color of the points represents the CNN output. Highlighted in black are the NS (ES) signals that have a CNN output of less (greater) than 0.5.}
\end{center}
\end{figure}

\begin{figure}
\begin{center}
\includegraphics[width=0.7\textwidth]{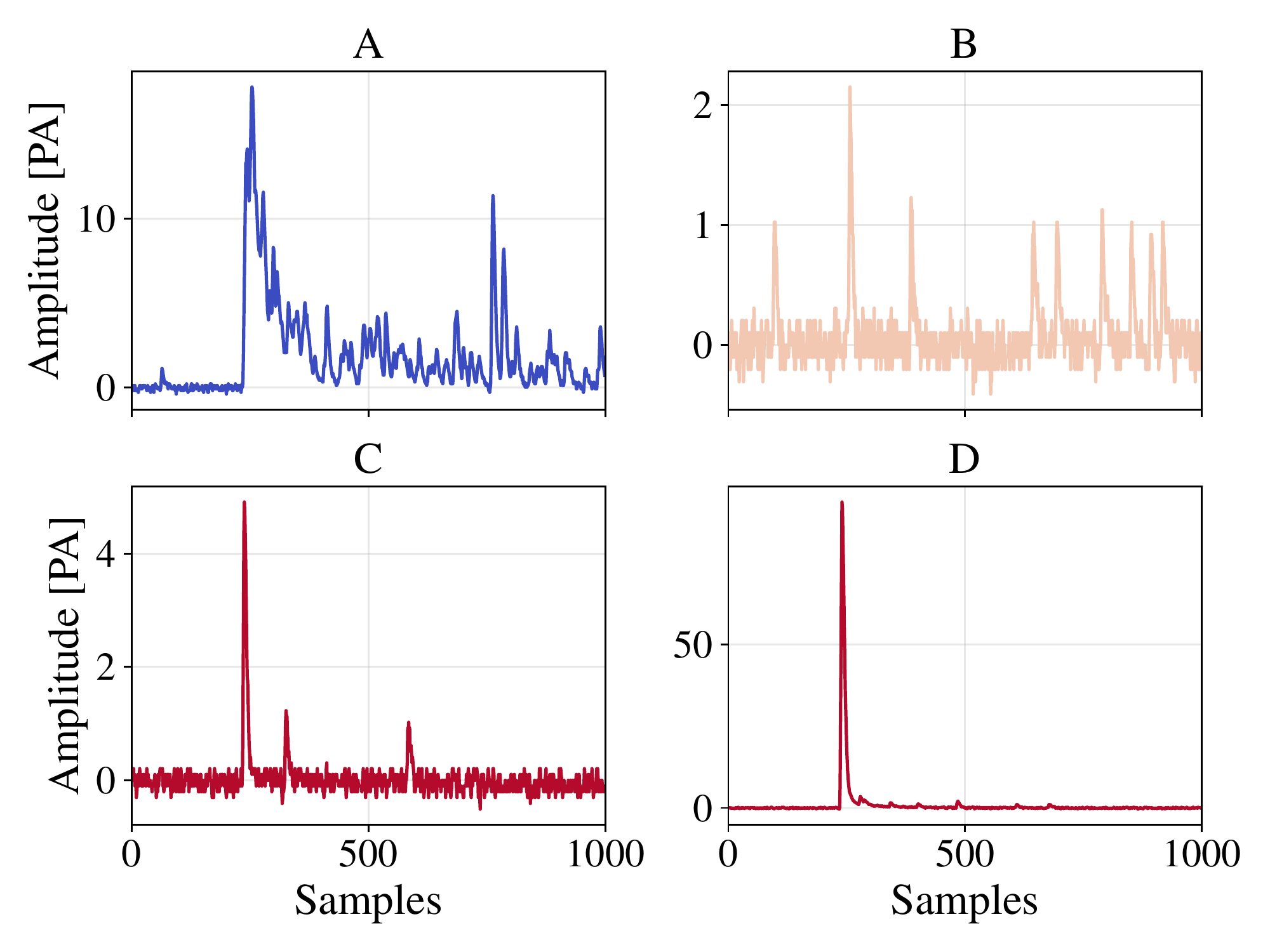}
\caption{\label{fig:tsne_samples} Representative signals for each of the labeled regions identified in the 2d t-SNE embedding. A: High energy NS signals. B: Pile up of single photon avalanche signals. C: Low energy ES signals. D: High energy muon signals.}
\end{center}
\end{figure}

Futher understanding of how the CNN interprets the signals can be obtained by visualising the high dimensional representation of each signal that is encoded within the 64-dimensional space of the fully-connected layer. The t-SNE algorithm \cite{tsne} is used to create a two-dimensional embedding of this space, which preserves a level of the local and global structure of the original space. This allows us to visualise in a qualitative way both the features encoded by the CNN, and to determine clusters of signals that have similar representations. The results are shown in Fig.~\ref{fig:tsne}, with each signal in this space coloured by the magnitude of the CNN output. It can be seen that the t-SNE shows the signals spread along an S-shape, forming two distinct ES and NS clusters. Within these clusters, four regions (A, B, C, D) of distinctly different signals have been highlighted, with representative signals in these regions shown in Fig.~\ref{fig:tsne_samples}. 
By examining these regions we find that there is is a continuous substructure of signals with varying energies, that gradually transforms from ES to NS signals, and that the CNN is using both the shape and energy information to discriminate between signals. The upper cluster consists entirely of NS signals, with those close to region A having a large amplitude pulse and a slowly decaying tail with many peaks, characteristic of high energy NS signals. Moving through this cluster we find waveforms with a continuous decrease in the amplitude and length of the NS pulse.

A similar structure is seen for the lower cluster of ES signals. The lower energy ES signals with a single prompt peak are found close to C, and in region D are high amplitude saturated signals, most likely caused by extremely energetic atmospheric muons. Region B contains the pulses that the CNN is not able to confidently classify as either ES or NS. These appear to be a collection of low energy signals that have several prompt peaks, most likely to be the pile-up of single photon avalanches. 

The signals highlighted in black are the NS (ES) signals that have a CNN output value of less (greater) than 0.5, and would therefore likely be misclassified when threshold cut is placed on the CNN output. The majority of these events appear in the cluster of ES signals and upon further inspection of these signals, they are clearly ES or pile-up signals that were incorrectly labelled by the simple selection shown in Figure 3. The fraction of misclassified signals is $\mathcal{O}(1\%)$ and is therefore in agreement with the original estimation of the contamination of the NS sample. This demonstrates that the CNN is robust to, at least, a small proportion of mislabeled training data and can correct the originally mislabelled signals. 

Furthermore, it is possible to use the discovered substructure shown in Fig.~\ref{fig:tsne_samples} to further sub-classify and divide pulses beyond that of the initial labeling of ES and NS, and could be used for example to remove pile-up or introduce a more suitable set of labels for the signals. To summarise, our investigation suggests that first, it is not necessary to have a perfect set of labels, the CNN is robust to a small proportion of misclassification and second, supervised learning is effective at recognising a wide range of sub-structure in the data. Through the exploration and clustering of data in this low-dimensional embedding, it is possible to further explore and understand a raw detector dataset without requiring unsupervised learning techniques. 

\section{Conclusions}
\label{conc}
In this work we have demonstrated a convolutional neural network architecture that provides superior performance in the classification of digitised signals compared to traditional PSD methods, achieving an AUC of $0.995\pm0.003$ on the set of signals obtained from a single \lith~PVT detector element. By investigating the representations learned by the CNN we have shown that it is possible to interpret the features extracted by the convolutional layers and therefore gain an understanding of how the CNN discriminates between different signal types. Further to this, we have shown that by visualising the high-dimensional representations it is possible to identify substructure within the signal types, even though the CNN was trained to perform a binary classification task. This approach could be utilised in many other areas of physics data analysis, such as to discover clusters of events in raw detector data without relying on hand-crafted variables.

\section*{Acknowledgments}
This work was supported by the Science \& Technology Facilities Council (STFC), United Kingdom and the European Research Council under the European Union's Horizon 2020 Programme (H2020-CoG)/ERC Grant Agreement \mbox{n. 682474} (corresponding author). We also thank S. Ihantola for developing the signal acquisition technique and providing the data sample used in this study. 



%% References


%\bibliographystyle{model1a-num-names}
\bibliographystyle{elsarticle-num}
\bibliography{biblio}

\begin{thebibliography}{10}
\expandafter\ifx\csname url\endcsname\relax
  \def\url#1{\texttt{#1}}\fi
\expandafter\ifx\csname urlprefix\endcsname\relax\def\urlprefix{URL }\fi
\expandafter\ifx\csname href\endcsname\relax
  \def\href#1#2{#2} \def\path#1{#1}\fi

\bibitem{SoLid}
Y.~Abreu, et~al., A novel segmented-scintillator antineutrino detector, JINST
  12~(04) (2017) P04024.
\newblock \href {http://dx.doi.org/10.1088/1748-0221/12/04/P04024}
  {\path{doi:10.1088/1748-0221/12/04/P04024}}.

\bibitem{LiFZnS.PSD}
E.~Legler, W.~Attwenger, F.~May, G.~Quittner, {Pulse Shape Discrimination
  System for $^6$LiF(ZnS) Scintillation Counters}, Review of Scientific
  Instruments 36 (1965) 1167.
\newblock \href {http://dx.doi.org/10.1063/1.1719829}
  {\path{doi:10.1063/1.1719829}}.

\bibitem{wavelet}
S.~Yousefi, L.~Lucchese, M.~Aspinall, Digital discrimination of neutrons and
  gamma-rays in liquid scintillators using wavelets, Nucl. Instrum. Meth.
  A598~(2) (2009) 551--555.
\newblock \href {http://dx.doi.org/10.1016/j.nima.2008.09.028}
  {\path{doi:10.1016/j.nima.2008.09.028}}.

\bibitem{Flores:2016dfj}
J.~L. Flores, I.~Martel, R.~Jiménez, J.~Galán, P.~Salmerón, {Application of
  neural networks to digital pulse shape analysis for an array of silicon strip
  detectors}, Nucl. Instrum. Meth. A830 (2016) 287--293.
\newblock \href {http://dx.doi.org/10.1016/j.nima.2016.05.107}
  {\path{doi:10.1016/j.nima.2016.05.107}}.

\bibitem{6551092}
T.~S. Sanderson, C.~D. Scott, M.~Flaska, J.~K. Polack, S.~A. Pozzi, Machine
  learning for digital pulse shape discrimination, in: 2012 IEEE Nuclear
  Science Symposium and Medical Imaging Conference Record (NSS/MIC), 2012, pp.
  199--202.
\newblock \href {http://dx.doi.org/10.1109/NSSMIC.2012.6551092}
  {\path{doi:10.1109/NSSMIC.2012.6551092}}.

\bibitem{deeplearning}
Y.~LeCun, Y.~Bengio, G.~Hinton, {Deep Learning}, Nature 521 (2015) 436.
\newblock \href {http://dx.doi.org/10.1038/nature14539}
  {\path{doi:10.1038/nature14539}}.

\bibitem{Komiske2017}
P.~T. Komiske, E.~M. Metodiev, M.~D. Schwartz, Deep learning in color: towards
  automated quark/gluon jet discrimination, JHEP 2017~(1) (2017) 110.
\newblock \href {http://dx.doi.org/10.1007/JHEP01(2017)110}
  {\path{doi:10.1007/JHEP01(2017)110}}.

\bibitem{next}
J.~Renner, et~al., Background rejection in next using deep neural networks,
  JINST 12~(01) (2017) T01004.
\newblock \href {http://dx.doi.org/10.1088/1748-0221/12/01/T01004}
  {\path{doi:10.1088/1748-0221/12/01/T01004}}.

\bibitem{panda}
H.~Qiao, C.~Lu, X.~Chen, K.~Han, X.~Ji, S.~Wang, Signal-background
  discrimination with convolutional neural networks in the {PandaX-III}
  experiment (2018).
\newblock \href {http://arxiv.org/abs/1802.03489} {\path{arXiv:1802.03489}}.

\bibitem{Delaquis}
S.~Delaquis, et~al., Deep neural networks for energy and position
  reconstruction in {EXO-200} (2018).
\newblock \href {http://arxiv.org/abs/1804.09641} {\path{arXiv:1804.09641}}.

\bibitem{DayaBay}
E.~Racah, et~al., {Revealing Fundamental Physics from the Daya Bay Neutrino
  Experiment using Deep Neural Networks} (2016).
\newblock \href {http://arxiv.org/abs/1601.07621} {\path{arXiv:1601.07621}}.

\bibitem{NOVA}
A.~Aurisano, et~al., {A Convolutional Neural Network Neutrino Event
  Classifier}, JINST 11~(09) (2016) P09001.
\newblock \href {http://dx.doi.org/10.1088/1748-0221/11/09/P09001}
  {\path{doi:10.1088/1748-0221/11/09/P09001}}.

\bibitem{miniboone}
R.~Acciarri, et~al., Convolutional neural networks applied to neutrino events
  in a liquid argon time projection chamber, JINST 12~(03) (2017) P03011.
\newblock \href {http://dx.doi.org/10.1088/1748-0221/12/03/P03011}
  {\path{doi:10.1088/1748-0221/12/03/P03011}}.

\bibitem{nn_review}
W.~Rawat, Z.~Wang, Deep convolutional neural networks for image classification:
  A comprehensive review, Neural Computation 29~(9) (2017) 2352--2449.
\newblock \href {http://dx.doi.org/10.1162/neco_a_00990}
  {\path{doi:10.1162/neco_a_00990}}.

\bibitem{chollet2015keras}
F.~Chollet, et~al., \href{https://keras.io}{Keras} (2015).
\newline\urlprefix\url{https://keras.io}

\bibitem{tensorflow2015-whitepaper}
M.~{Abadi}, et~al., {TensorFlow: A system for large-scale machine learning}
  (2016).
\newblock \href {http://arxiv.org/abs/1605.08695} {\path{arXiv:1605.08695}}.

\bibitem{adam}
D.~P. Kingma, J.~Ba, Adam: {A} method for stochastic optimization, CoRR
  abs/1412.6980.
\newblock \href {http://arxiv.org/abs/1412.6980} {\path{arXiv:1412.6980}}.

\bibitem{tsne}
L.~van~der Maaten, G.~Hinton, Visualizing data using {t-SNE}, Journal of
  Machine Learning Research 9 (2008) 2579--2605.

\end{thebibliography}

%% not want to use model3-num-names.bst.





% \end{thebibliography}

\end{document}